\documentclass{article}

\usepackage[english]{babel}

\usepackage[letterpaper,top=2cm,bottom=2cm,left=3cm,right=3cm,marginparwidth=1.75cm]{geometry}

\usepackage{amsmath}
\usepackage{graphicx}
\usepackage[colorlinks=true, allcolors=blue]{hyperref}
\usepackage{amsfonts}
\usepackage{mathtools}

\newcommand{\NN}{\mathbb{N}}

\newcommand\xor{\mathbin{\char`\^}}
\renewcommand\xor{\mathbin{^\wedge}}

\newcommand{\uand}{\mathbin{\&}}
\newcommand{\uor}{\mathbin{|}}
\newcommand\uxor{\xor}
\newcommand\unot{\mathord{\sim}}

\usepackage{float}
\usepackage{kbordermatrix}
\renewcommand{\kbldelim}{(}
\renewcommand{\kbrdelim}{)}

\usepackage{multirow}
\usepackage{tikz}

\usepackage{colonequals}

\title{Deobfuscation of Semi-Linear Mixed Boolean-Arithmetic Expressions}
\author{Colton Skees \\ Mazeworks Security \\ \href{colton@mazeworkssecurity.com}{colton@mazeworkssecurity.com} }

\begin{document}
\maketitle

\begin{abstract}
Mixed Boolean-Arithmetic (MBA) obfuscation is a common technique used to transform simple expressions into semantically equivalent but more complex combinations of boolean and arithmetic operators. Its widespread usage in DRM systems, malware, and software protectors is well documented.

In 2021, Liu et al. proposed a groundbreaking method of simplifying linear MBAs, utilizing a hidden two-way transformation between 1-bit and n-bit variables. In 2022, Reichenwallner et al. proposed a similar but more effective method of simplifying linear MBAs, SiMBA, relying on a similar but more involved theorem. However, because current linear MBA simplifiers operate in 1-bit space, they cannot handle expressions which utilize constants inside of their bitwise operands, e.g. (x\&1), (x\&1111) + (y\&1111).

We propose an extension to SiMBA that enables simplification of this broader class of expressions. It surpasses peer tools, achieving efficient simplification of a class of MBAs that current simplifiers struggle with.

\end{abstract}

\section{Introduction}

Mixed Boolean-Arithmetic (MBA) obfuscation is a popular technique for code obfuscation, originally introduced in 2006 by Zhou et al. This technique allows simple expressions to be represented as a semantically equivalent but more complex combination of boolean and arithmetic operators, thereby making it harder for an attacker to understand or recover the original expression. 

A key strength of MBA obfuscation lies in its resilience against traditional analysis tools. Tools designed solely for arithmetic or boolean expressions (e.g. SAT Solvers, Boolean Minimizers, Computer Algebra Systems) cannot handle the mixing of boolean and arithmetic operators. Until recently, no publicly known methods existed for effectively analyzing or simplifying MBA expressions\cite{Liu_2021}		

However, recent research has yielded progress in tackling MBAs. Tools such as MBA-Blast, MBA-Solver, and SiMBA allow efficient simplification of a sub-class of MBAs known as linear MBAs. Tools such as GAMBA\cite{Reichenwallner_2023} and ProMBA\cite{Lee2023} expand on this research further, combining linear MBA simplifiers with ad hoc techniques to achieve simplification of general MBAs.

Still, current MBA simplifiers exhibit a critical limitation: they cannot effectively simplify expressions which utilize constants inside of their bitwise operators. For instance, $((22079729|(5368709120\uand x))+(5368709207\uxor (5368709120\uand x))) - 5390788936$ $\pmod{2^{64}}$ is equivalent to zero, but it cannot be simplified soundly by current nonlinear MBA solvers unless an SMT solver is used.

It is well documented that MBAs with nontrivial bitwise constants are among the most challenging MBAs to simplify\cite{Reichenwallner_2023}. Motivated by these limitations, we contribute an extension to SiMBA which enables simplification of expressions with nontrivial bitwise constants. In addition to supporting a broader class of expressions, it is very competitive in runtime and inherits many benefits of SiMBA - namely that it does not require a decomposition of an MBA into a canonical input format, and the implementation is independent of the number of variables and the MBAs’ complexity.

\section{Preliminaries}

\subsection{Linear Mixed Boolean-Arithmetic Expressions}
Quoting verbatim from SiMBA\cite{Reichenwallner_2022},

	Let $B = \{0,1\}$ and $n,t \in \NN$. A \textit{linear mixed Boolean-arithmetic expression (MBA)} with values in $B^n$ and $t$ variables is a function $e: \left(B^n\right)^t \to B^n$ of the form $$e\left(x_1,\ldots,x_t\right) = \sum_{i\in I} a_i e_i\left(x_1,\ldots,x_t\right),$$ where $I \subset \NN$ is an index set, $a_i \in B^n$ are constants and $e_i$ are bitwise expressions of $x_1,\ldots,x_t$ for $i \in I$.

$\newline$

Notably, the bitwise expressions may not contain \textit{nontrivial} constants, i.e.,  constants other than 0 or -1.

\subsection{N-bit to 1 bit transformation}

In 2007, Zhou et al proved that Linear MBAs are equivalent in B$^n$ if they are equivalent on $B = \{0, 1\}$.

It wasn't until 2021 that Liu et al. pointed out that this theorem could be used in the reverse direction, allowing simplification of linear MBAs living in B$^n$ via a transformation to 1-bit space\cite{Liu_2021}. This finding was stated in Zhou et al.'s original paper, but remained unnoticed due to a mistake \cite{Reichenwallner_2022}

Reichenwallner et al. took this a step further, proving that a full transformation to 1-bit space is not required. Instead, they evaluate whole linear MBAs for all combinations of zeroes and ones, yielding vectors living in B$^n$.

\subsection{Semi-linear Mixed Boolean-Arithmetic Expressions}
A semi-linear MBA expression is an MBA that would be linear, if not for nontrivial constants within the bitwise subexpressions. The term was first introduced by Sok et al in SSLEM\cite{mok2022sslem}.

Unlike linear MBAs, semi-linear MBAs are not guaranteed to be equivalent in B$^n$ if they are equivalent on $B = \{0, 1\}$. Consequently, the N-bit to 1 bit transform is not applicable to semi-linear MBAs, thereby making linear MBA simplifiers incompatible with semi-linear MBAs.

\subsection{N-bit to N-bit transformation}
The N-bit to 1-bit transform can be extended to work on semi-linear MBAs.

\subsubsection{Theorem 1.0: Decomposing linear and semi-linear expressions }
We want to show that any linear or semi-linear expression can be decomposed into a linear combination of disjoint bitwise expressions, with each bitwise term being applied to only a single bit. We can do so using the following steps:

\begin{enumerate}
  \item \textit{Converting univariate terms into bitwise terms}: Given a function $e(x0, x1, x...)$ with $t$ variables, a univariate term such as $x$ can be decomposed into a bitwise expression, via construction of a boolean truth table with $2^t$ entries - where each entry is set to $true$ only when $x$ is $true$. As an example, $x$ is equivalent to $(x \uand y)\uor (x\uand \unot y)$ in the bivariate case.
  \item \textit{Bit-blasting bitwise terms}: Because bitwise operations are applied to each bit separately, they can be decomposed into linear combinations of bitwise terms over individual bits. As an example, $2 * (x\uxor y)$ $\pmod{2^3}$ can be decomposed into $2 * (((x\uxor y)\uand 1) \uor  ((x\uxor y)\uand 2) \uor  ((x\uxor y)\uand 4))$, and subsequently $2 * (((x\uxor y)\uand 1) + ((x\uxor y)\uand 2) + ((x\uxor y)\uand 4))$ after applying the identity $(x\uor y) == (x+y)$ if $x$ and $y$ are disjoint.
  \item \textit{Decomposing bitwise terms:} A boolean function can be decomposed into a linear combination of disjoint base bitwise expressions, utilizing the truth table of a boolean function or other means. Using the earlier example, $(x \uxor y)$ can be decomposed into $(x\uand \unot y)+(\unot x\uand y)$. Applying this transformation to our earlier example yields $2 * ((((x\uand \unot y)+(\unot x\uand y))\uand 1) + (((x\uand \unot y)+(\unot x\uand y))\uand 2) + (((x\uand \unot y)+(\unot x\uand y))\uand 4))$

  This decomposition has the useful property that when one base bitwise expression evaluates to $true$, all other bitwise expressions evaluate to false - making them independent of one another.

  \item \textit{Distributing:} Using the identity from (1), the bitwise constant masks can be distributed into each base bitwise expression. Considering our earlier example, $(((x\uand \unot y)+(\unot x\uand y))\uand 4)$ can be transformed into $(((x\uand \unot y)\uor(\unot x\uand y))\uand 4)$, then subsequently $((x\uand \unot y)\uand 4)\uor ((\unot x\uand y)\uand 4)$ using the identity $x\uand (y\uor z)$ == $(x\uand y)\uor (x\uand z)$ - finally arriving at $((x\uand \unot y)\uand 4)+((\unot x\uand y)\uand 4)$ after applying the identity from (1) again. Applying this transformation to our full example yields $2 * ((((x\uand \unot y)\uand 1)+((\unot x\uand y)\uand 1)) + (((x\uand \unot y)\uand 2)+((\unot x\uand y)\uand 2)) + (((x\uand \unot y)\uand 4)+((\unot x\uand y)\uand 4)))$. Multiplication can then be distributed over the linear combination.
  
\end{enumerate}$\newline$

The final result of this decomposition on our earlier example is $2*(((x\uand \unot y)\uand 1))+2*(((\unot x\uand y)\uand 1))+2*(((x\uand \unot y)\uand 2))+2*(((\unot x\uand y)\uand 2))+2*(((x\uand \unot y)\uand 4))+2*(((\unot x\uand y)\uand 4))$. 

\subsubsection{Theorem 1.1: Construction of semi-linear signature vectors and equivalence on $B = \{0, 2^i\}$ }
Knowing that a vector can be represented as a linear combination of a set of base vectors, we can represent linear and semi-linear expressions as a vector, where the base vectors are truth table representations of bitwise expressions. In particular, our base vectors are truth table representations of all $2^t$ possible truth values, where $t$ is the number of input variables. 

This representation is referred to as a \textit{signature vector}, and was first introduced in MBA-Blast \cite{Liu_2021}. Linear MBAs are equivalent if their signature vectors coincide, and as stated in \cite{Reichenwallner_2022}, for linear MBAs a signature vector can be constructed via a direct evaluation of the input expression for all possible truth values, yielding a vector living in B$^n$.

Constructing a signature vector for a \textit{semi-linear} expression can be done using a similar but more involved method. Knowing that the decomposition in Theoreom 1.0 is possible for any semi-linear expression, a vector can be setup in the form:

$
   \mathbf{e(x, y)} = 
     \bordermatrix{ & \unot x \uand \unot y & x \uand \unot y & \unot x \uand y & x \uand y \cr
       Bit 0 & ? & ? & ? & ? \cr
       Bit 1& ? & ? & ? & ?\cr
       Bit 2& ? & ? & ? & ?\cr
       ... & ? & ? & ? & ?\cr
       } \qquad
$

$\newline$

As an example, we consider the simple semi-linear MBA $e(x, y) = (x\&5) + (y\&3)$ $\pmod{2^3}$

$\newline$

To find the coefficients, we must evaluate the expression on $B = \{0, 2^i\}$ for each bit $i \in \{0 ... N\}$. This yields three vectors:

$
   \mathbf{Bit0(x, y)} = 
     \bordermatrix{ &  \cr
       e(0, 0) & 0\cr
       e(1, 0) & 1\cr
       e(0, 1) & 1\cr
       e(1, 1) & 2\cr
       } \qquad
$
$
   \mathbf{Bit1(x, y)} = 
     \bordermatrix{ &  \cr
       e(0, 0) & 0\cr
       e(2, 0) & 0\cr
       e(0, 2) & 2\cr
       e(2, 2) & 2\cr
       } \qquad
$
$
   \mathbf{Bit2(x, y)} = 
     \bordermatrix{ &  \cr
       e(0, 0) & 0\cr
       e(4, 0) & 4\cr
       e(0, 4) & 0\cr
       e(4, 4) & 4\cr
       } \qquad
$

$\newline$

, then: $
   \mathbf{e(x, y)} = 
     \bordermatrix{ & \unot x \uand \unot y & x \uand \unot y & \unot x \uand y & x \uand y \cr
       Bit 0 & 0 & 1 & 1 & 2 \cr
       Bit 1& 0 & 0 & 2 & 2\cr
       Bit 2& 0 & 4 & 0 & 4\cr
       } \qquad
$

$\newline$

Finally, each evaluation result must be shifted down by $i$ accordingly. Since multiplication can only preserve or increase the number of trailing zero bits, shifting by $i$ is guaranteed to not discard information, and is therefore correct. When removing the constant offset corresponding to the 0th vector entry, as done in later sections, we shift the constant offset down by $i$ before subtraction.

Applying the shift to the result vector yields: 

$
   \mathbf{e(x, y)} = 
     \bordermatrix{ & \unot x \uand \unot y & x \uand \unot y & \unot x \uand y & x \uand y \cr
       Bit 0 >> 0 & 0 & 1 & 1 & 2 \cr
       Bit 1 >> 1& 0 & 0 & 1 & 1\cr
       Bit 2 >> 2& 0 & 1 & 0 & 1\cr
       } \qquad
$,

$\newline$

which can be decomposed into a linear combination over individual bits,
yielding a solution: $e(x, y) = \newline 0*(1\uand (\unot x\uand \unot y)) + 1*(1\uand (x\uand \unot y)) + 1*(1\uand (\unot x\uand y)) + 2*(1\uand (x\uand y)) + \newline
0 * (1 \uand (\unot x \uand \unot y)) + 1 * (1 \uand (x \uand \unot y)) + 1 * (1 \uand (\unot x \uand y)) + 2 * (1 \uand (x \uand y)) +\newline
0 * (2 \uand (\unot x \uand \unot y)) + 0 * (2 \uand (x \uand \unot y)) + 1 * (2 \uand (\unot x \uand y)) + 1 * (2 \uand (x \uand y)) +\newline
0 * (4 \uand (\unot x \uand \unot y)) + 1 * (4 \uand (x \uand \unot y)) + 0 * (4 \uand (\unot x \uand y)) + 1 * (4 \uand (x \uand y))
$

Since a semi-linear result vector can be seen as a canonical representation of a semi-linear MBA, and the construction of this vector is independent of the complexity and structure of its input, two semi-linear expressions are equivalent if their signature vectors coincide. 

$\newline$

SSLEM\cite{mok2022sslem} introduced a similar decomposition, however they did not delve deeply into it.

\section{Simplification of Linear Mixed Boolean-Arithmetic Expressions using SiMBA}
Many tools exist for simplifying linear MBAs, namely SiMBA, MBA-Blast, and MBA-Solver. Because our algorithm is based off of SiMBA, and because SiMBA surpasses all peer tools, we omit the implementation details of MBA-Blast and MBA-Solver and turn our attention to SiMBA.

For our extension, we reimplemented SiMBA in 5000 lines of C\#. Our solution also implements the extensions proposed in \cite{Reichenwallner_2023}, allowing SiMBA to find simpler solutions. The source code is available at \href{https://github.com/mazeworks-security/MSiMBA}{https://github.com/mazeworks-security/MSiMBA}.

\subsection{Computing the signature vector}
SiMBA first calculates a signature vector for a whole linear MBA. As an example, we consider the MBA $e(x, y) = x+y$ $\pmod{2^{64}}$

$\newline$

Constructing a signature vector yields: 

$\newline$

$
   \mathbf{e(x, y)} = 
     \bordermatrix{ & \unot x \uand \unot y & x \uand \unot y & \unot x \uand y & x \uand y \cr
       & 0 & 1 & 1 & 2 \cr
       } \qquad
$

$\newline$

Yielding a solution: $e(x, y) = 0*(x\uand \unot y) + 1*(x\uand \unot y) + 1*(\unot x\uand y) + 2*(x\uand y)$

$\newline$
$\newline$

Turning our attention to another example: $e(x, y) = \unot(x+y)$ has the vector

$\newline$

$
   \mathbf{e(x, y)} = 
     \bordermatrix{ & \unot x \uand \unot y & x \uand \unot y & \unot x \uand y & x \uand y \cr
       & -1 & -2 & -2 & -3 \cr
       } \qquad
$

$\newline$

In order to find a valid solution, the constant offset corresponding to the first entry must be subtracted from all other vector entries, yielding:

$\newline$

$
   \mathbf{e(x, y)} = 
     \bordermatrix{ & \unot x \uand \unot y & x \uand \unot y & \unot x \uand y & x \uand y \cr
       & 0 & -1 & -1 & -2 \cr
       } \qquad
$

$\newline$

Finally, arriving at a correct but suboptimal solution:

$\newline$

$-1 + -1*(x\uand \unot y) + -1*(\unot x\uand y) + -2*(x\uand y)$

$\newline$

, after prepending the constant offset.

\subsection{Finding a linear combination}

From earlier examples, it's clear that decomposing an expression into $2^N$ terms (one term for each base bitwise expression) will not always yield good results. Especially in cases where the ground truth contains no bitwise terms, e.g. 'x + y', this falls short.

To address this, SiMBA derives a linear combination in a generic way that prefers arithmetic terms over mixed or boolean terms. The specific implementation details are not revisited here, but can be found in \cite{Reichenwallner_2022}.

\subsection{Finding a simpler solution}

In order not to miss very simple solutions, SiMBA performs several refinement attempts. The original refinement attempts are not revisited in this paper.

In GAMBA, Reichenwallner et al. extended SiMBA with additional refinement attempts - namely variable partitioning, identification of simple negations, and boolean minimization using the Quine-McCluskey algorithm\cite{Reichenwallner_2023}. In our implementation, we extended SiMBA to use Espresso for boolean minimization, and integrated the four variable truth table for optimal simplification of four variable boolean functions. 

\section{Our Extension: Simplification of Semi-Linear Mixed Boolean-Arithmetic expressions}

\subsection{Computing the signature vector}
In order to capture the full semantics of a semi-linear MBA, a semi-linear signature vector must be built using theorem 1.1.
As an example, we consider the MBA $e(x, y) = 2*(x\&5) + 2*(y\&3)$ $\pmod{2^3}$

$\newline$

We decompose the expression into the vector:

$\newline$

$
   \mathbf{e(x, y)} = 
     \bordermatrix{ & \unot x \uand \unot y & x \uand \unot y & \unot x \uand y & x \uand y \cr
       Bit 0 & 0 & 2 & 2 & 4 \cr
       Bit 1& 0 & 0 & 2 & 2\cr
       Bit 2& 0 & 2 & 0 & 2\cr
       } \qquad
$

$\newline$

\subsection{Finding a linear combination}
After subtracting the constant offset from all signature vector entries, 
we apply SiMBA's logic to find an initial linear combination for each bit's signature vector row. Using the vector from above, we arrive at a solution of: $\newline\newline$
$2*(x\uand 1) + 2*(y\uand 1) + \newline
2*(y\uand 2) + \newline
2*(x\uand 4)$

$\newline$
This is valid in any case but far from optimal, because the resulting linear combination is effectively computing each bit of the base bitwise expressions individually. Fortunately the results can be drastically improved using a refinement procedure - for which the exact ground truth is recovered on our current example. In the next section we detail this refinement procedure.

\section{Refining a semi-linear solution}

\subsection{Preliminaries}
Semi-linear MBA simplification introduces a new set of challenges that are not present with linear MBA simplification, stemming from the fact that there is a new degree of flexibility in both the coefficients and masks being applied over bitwise expressions. As an example, the 8-bit MBA $64*(130\uand x)$ is equivalent to both $192*(2\uand x)$ and $64*(2\uand x)$, despite the differences between their coefficients and masks. 

This is particularly problematic when trying to refine a solution via merging of terms. We consider the 8-bit MBA: $64*(130\uand x) + 64*(192\uand x)$, which can be simplified to $64*(194\uand x)$ only after rewriting $64*(130\uand x)$ to $64*(2\uand x)$, then applying the identity: $(m1*(c1\&x) + m1*(c2\&x)) == m1*((c1\uor c2)\uand x)$ if $c1$ and $c2$ are disjoint. 

In order to perform the refinement from above, aswell as the refinements listed below, we need to be able to determine whether a term's coefficient or bitmask can be changed. While the earlier examples are trivially solvable, there are an arbitrary number of similar instances of this problem - and to the best of our knowledge there is no general ruleset that can be applied to solve this.

Below we list two methods, $CanChangeCoefficientTo$ and $CanChangeMaskTo$, which perform a comparison of signature vectors to determine whether a term's coefficient or bitmask can be changed. Our refinement procedure relies heavily on the application of these methods, aswell as similar variations of them.
\begin{verbatim}
// Returns true if an expression in the form of `m1 * (mask&x)` can be changed to 
// m2*(mask&x)`.
bool CanChangeCoefficientTo(uint bitWidth, ApInt oldCoeff, ApInt newCoeff, 
ApInt bitMask, ApInt moduloMask)
{
    for (ushort i = 0; i < bitWidth; i++)
    {
        var value = (ApInt)1 << i;
        var op1 = moduloMask & (oldCoeff * (value & bitMask));
        var op2 = moduloMask & (newCoeff * (value & bitMask));
        if (op1 != op2)
            return false;
    }

    return true;
}
\end{verbatim}

$\newline$
$\newline$

\begin{verbatim}
// Returns true if an expression in the form of `m1 * (oldMask&x)` can be changed to 
// m1*(newMask&x)`.
bool CanChangeMaskTo(uint bitWidth, ApInt coeff, ApInt oldMask, 
ApInt newMask, ApInt moduloMask)
{
    for (ushort i = 0; i < bitWidth; i++)
    {
        var value = (ApInt)1 << i;
        var op1 = moduloMask & (coeff * (value & oldMask));
        var op2 = moduloMask & (coeff * (value & newMask));
        if (op1 != op2)
            return false;
    }

    return true;
}
\end{verbatim}

For the remainder of this section, all expressions should be assumed to be 64-bit.

\subsection{Reducing the number of terms corresponding to each basis expression}
The refinement procedure begins by trying to reduce the number of terms corresponding to each base bitwise expression. We apply the following refinements:

\begin{enumerate}
  \item If two terms with disjoint bitmasks exist in the form of $(m1*(c1\&x) + m1*(c2\&x))$, collapse them into a single term using the identity: $(m1*(c1\&x) + m1*(c2\&x)) == m1*((c1\uor c2)\uand x)$ if $c1$ and $c2$ are disjoint.

  \item If two terms with disjoint bitmasks exist in the form of $(m1*(c1\&x) + m2*(c2\&x))$, and the second coefficient can be changed to the first term's coefficient, collapse them into a single term using the identity in (1).
  \item If a term's coefficient can be changed to zero, discard it.
  \item If a term's coefficient can be replaced with -1, change it.
  \item Repeat step 2.
  \item If there are three terms in the form of $(m1*(c1\&x) + m2*(c2\&x)) + m3*(c3\&x))$, the first two coefficients sum up to the third coefficient, and all three bitmasks are disjoint, express three terms as the summation of two new terms.

  Example: $e(x) = (529682\uand x) + 7676756576*(23429673\uand x) + 7676756577*(24772\uand x)$

  is transformed to $((529682|24772)\uand x) + 7676756576*((23429673|24772)\uand x)$,

  then finally constant folded to $(554454\uand x) + 7676756576*(23454445\uand x)$.
\end{enumerate}

\subsection{Moving away from a linear combination of conjunctions}
In hopes of finding a better solution, it is necessary to try to decompose the initial linear combination of conjunctions into a simpler linear combinations of arbitrary bitwise expressions. In this step of the refinement procedure, we try to recover high level structure(XORs), and attempt to minimize the AST size of the entire solution.

We apply the following refinements:
\begin{enumerate}
  \item If two terms with disjoint bitmasks exist in the form of $(m1*(c1\&x) + m2*(c2\&x))$, m2 is equivalent to $-1 * m1$, and $c2$ can be changed to a negation of $c1$, collapse them into a single XOR term.

  Example: $e(x, y) = 980+(-10*(98\uand x))+(10*(-99\uand x))$

  Apply the identity $m1*(x\uxor y) = m1*y + m1*(x\uand \unot y) - m1*(x\uand y)$,

  arriving at $10 * ((98\uxor x))$

  \item If two terms exist in the form of $(m1*(c1\&x) + m2*(c2\&x))$, try to express them as a linear combination of two terms where only one term has a bitmask.

  Example: $e(x) = 7*(1111\uand x) + 2*(-1112\uand x)$

  Apply the logic from section 5.1 to determine that the expression can be expressed as $(m1 - m2)*(c1\uand x) + m2*x$,

  arriving at $5*(1111\uand x) + 2*x$

  \item Repeat step (1)

  \item If multiple terms with the same basis expression exist, group them into a linear combination and invoke our extended version of SiMBA. If SiMBA returns a simpler result, keep it.

  \item Repeat step (1)
  
\end{enumerate}$\newline$

\subsection{Finding a better solution in 1-bit space}
As a last attempt at finding a better solution, we try to find a simpler solution in 1-bit space using SiMBA and substitution of bitwise constants.

Before running SiMBA, we try to minimize the number of variables by expressing say three bitwise constants as a boolean combination of two other bitwise constants. Consider the MBA: 

$e(x, y) = ((1111\uand (x\uand (\unot y)))|(2222\uand ((\unot x)\uand y)))|(3327\uand (x\uand y))$,

$\newline$

which can be expressed as $((1111\uand (x\uand (\unot y)))|(2222\uand ((\unot x)\uand y)))|((1111\uor 2222)\uand (x\uand y))$,

$\newline$

then $((c1\uand (x\uand (\unot y)))|(c2\uand ((\unot x)\uand y)))|((c1\uor c2)\uand (x\uand y))$ after substitution,

$\newline$

and finally $(x\uand c1)|(y\uand c2)$ after applying SiMBA.

$\newline$

After applying SiMBA and back substitution, we pick from the two solutions depending upon a cost function. 

\section{Taking a shortcut}
For a semi-linear linear MBA, one still may be able to find a solution in 1-bit space using the original SiMBA algorithm instead. As an example, $(x\uand 1111) + (x\uand -1112)$ is syntactically semi-linear, but its ground truth, $X$, is linear.

A simple method exists for determining whether a semi-linear MBA can be simplified in one-bit space:

\begin{enumerate}
  \item Construct a semi-linear signature vector for the MBA
  \item Subtract the constant offset from all vector entries
  \item Construct an expression corresponding to the behavior on $B = \{0, 1\}$
  \item Construct a semi-linear signature vector for the newly created expression
  \item Compare both semi-linear signature vectors
\end{enumerate}$\newline$

If the signature vectors are equivalent, then a solution can be found using linear MBA simplification techniques. Otherwise a solution must be found using using semi-linear simplification techniques.

\section{Verification and Comparison}
In the following, we evaluate our approach against two state of the art general MBA simplifiers: GAMBA and ProMBA. Past research indicates that Arybo, SSPAM, and Syntia do not reliably simplify MBAs in general\cite{Reichenwallner_2022}, hence we are content with only considering these two simplifiers.

All evaluations were performed on a windows machine with an i9-10850k CPU and 32 GiB of memory. For GAMBA, $PyPy$ was used as a runtime due to its superior performance on our benchmarks.

\subsection{Datasets}
In order to evaluate behavior for semi-linear expressions, we needed a large dataset of MBAs with bitwise constants. To construct this dataset, we substitute some variables with constants for the datasets provided by the repositories of SiMBA, and MBA-Obfuscator\cite{Liu2021SoftwareOW}.

In addition to modifying existing datasets, we contribute a self-generated dataset of semi-linear MBAs.

\subsection{Comparison on modified SiMBA datasets}

\begin{table}[H]
\centering
\begin{tabular}{|ccccccc|}
\hline
\multicolumn{7}{|c|}{\bfseries ProMBA}                                                                                                                                                                                                       \\ \hline
\multicolumn{1}{|l|}{\multirow{2}{*}{\bfseries Expr}} & \multicolumn{2}{c|}{\bfseries 2 variables}                                  & \multicolumn{2}{c|}{\bfseries 3 variables}                                  & \multicolumn{2}{c|}{\bfseries 4 variables}             \\ \cline{2-7} 
\multicolumn{1}{|l|}{}                      & \multicolumn{1}{c|}{\bfseries Runtime}   & \multicolumn{1}{c|}{\bfseries \# of Nodes} & \multicolumn{1}{c|}{\bfseries Runtime}   & \multicolumn{1}{c|}{\bfseries \# of Nodes} & \multicolumn{1}{c|}{\bfseries Runtime}   &\bfseries  \# of Nodes \\ \hline
\multicolumn{1}{|c|}{$e_1$}                    & \multicolumn{1}{c|}{34,722 $ms$} & \multicolumn{1}{c|}{16.66 / 3}       & \multicolumn{1}{c|}{33,687 $ms$} & \multicolumn{1}{c|}{60.61 / 3}       & \multicolumn{1}{c|}{31,984 $ms$} & 319.3 / 3       \\ \hline
\multicolumn{1}{|c|}{$e_2$}                    & \multicolumn{1}{c|}{32,015 $ms$} & \multicolumn{1}{c|}{14.75 / 1}       & \multicolumn{1}{c|}{37,452 $ms$} & \multicolumn{1}{c|}{53.13 / 1}       & \multicolumn{1}{c|}{31,799 $ms$} & 310.77 / 1       \\ \hline
\multicolumn{1}{|c|}{$e_3$}                    & \multicolumn{1}{c|}{32,379 $ms$} & \multicolumn{1}{c|}{17.68 / 5}       & \multicolumn{1}{c|}{36,813 $ms$} & \multicolumn{1}{c|}{67.54 / 5}       & \multicolumn{1}{c|}{32,221 $ms$} & 318.7 / 5       \\ \hline
\multicolumn{1}{|c|}{$e_4$}                    & \multicolumn{1}{c|}{42,419 $ms$} & \multicolumn{1}{c|}{19.66 / 7}       & \multicolumn{1}{c|}{42,616 $ms$} & \multicolumn{1}{c|}{65.93 / 7}         & \multicolumn{1}{c|}{31,687 $ms$} & 323.02 / 7       \\ \hline
\multicolumn{1}{|c|}{$e_5$}                    & \multicolumn{1}{c|}{37,852 $ms$} & \multicolumn{1}{c|}{17.84 / 4}       & \multicolumn{1}{c|}{37,644 $ms$} & \multicolumn{1}{c|}{58.75 / 4}       & \multicolumn{1}{c|}{31,923 $ms$} & 313.55 / 4       \\ \hline
\end{tabular}
\caption{Runtime and distance to ground truth of ProMBA on 1000 MBAs generated for five expressions}\label{tab:our}
\end{table}

In table one we see that ProMBA fails to simplify semi-linear MBAs in general, regardless of the number of variables. A large part of this can be explained by the fact that ProMBA does not have sufficient term rewriting rules for functions with nontrivial constants, and the underlying linear simplifier cannot be used to semi-linear expressions. Instead, ProMBA must resort to synthesis and therefore verification via SMT solving, which cannot be expected to terminate in a reasonable time on complex inputs.
\begin{table}[H]
\centering
\begin{tabular}{|ccccccc|}
\hline
\multicolumn{7}{|c|}{\bfseries GAMBA}                                                                                                                                                                                                       \\ \hline
\multicolumn{1}{|l|}{\multirow{2}{*}{\bfseries Expr}} & \multicolumn{2}{c|}{\bfseries 2 variables}                                  & \multicolumn{2}{c|}{\bfseries 3 variables}                                  & \multicolumn{2}{c|}{\bfseries 4 variables}             \\ \cline{2-7} 
\multicolumn{1}{|l|}{}                      & \multicolumn{1}{c|}{\bfseries Runtime}   & \multicolumn{1}{c|}{\bfseries \# of Nodes} & \multicolumn{1}{c|}{\bfseries Runtime}   & \multicolumn{1}{c|}{\bfseries \# of Nodes} & \multicolumn{1}{c|}{\bfseries Runtime}   &\bfseries  \# of Nodes \\ \hline
\multicolumn{1}{|c|}{$e_1$}                    & \multicolumn{1}{c|}{14.76 $ms$} & \multicolumn{1}{c|}{3.73 / 3}       & \multicolumn{1}{c|}{492.68 $ms$} & \multicolumn{1}{c|}{25.09 / 3}       & \multicolumn{1}{c|}{9517.67 $ms$} & 304.77 / 3       \\ \hline
\multicolumn{1}{|c|}{$e_2$}                    & \multicolumn{1}{c|}{3.9 $ms$} & \multicolumn{1}{c|}{1.07 / 1}       & \multicolumn{1}{c|}{1256.3 $ms$} & \multicolumn{1}{c|}{12.41 / 1}       & \multicolumn{1}{c|}{8592.48 $ms$} & 295.03 / 1       \\ \hline
\multicolumn{1}{|c|}{$e_3$}                    & \multicolumn{1}{c|}{10.08 $ms$} & \multicolumn{1}{c|}{6.26 / 5}       & \multicolumn{1}{c|}{513.1 $ms$} & \multicolumn{1}{c|}{27.58 / 5}       & \multicolumn{1}{c|}{9406.09 $ms$} & 310.05 / 5       \\ \hline
\multicolumn{1}{|c|}{$e_4$}                    & \multicolumn{1}{c|}{18.38 $ms$} & \multicolumn{1}{c|}{11.1 / 7}       & \multicolumn{1}{c|}{593.27 $ms$} & \multicolumn{1}{c|}{32.3 / 7}         & \multicolumn{1}{c|}{9570.36 $ms$} & 308.69 / 7       \\ \hline
\multicolumn{1}{|c|}{$e_5$}                    & \multicolumn{1}{c|}{8.27 $ms$} & \multicolumn{1}{c|}{4.77 / 4}       & \multicolumn{1}{c|}{375.83 $ms$} & \multicolumn{1}{c|}{20.93 / 4}       & \multicolumn{1}{c|}{9355.84 $ms$} & 304.33 / 4       \\ \hline
\end{tabular}
\caption{Runtime and distance to ground truth of GAMBA on 1000 MBAs generated for five expressions}\label{tab:our}
\end{table}

In table two we see that GAMBA performs well on expressions with a small number of variables and correspondingly bitwise constants - simplifying most two variable expressions closely to their ground truth. As the number of variables and bitwise constants grows, GAMBA's runtime and output quality degrades exponentially. 
This can largely be explained by the exponential runtime of linear MBA simplification. As the number of bitwise constants grows, GAMBA must substitute the constants with an increasingly large number of temporary variables before attempting linear subtree simplification. Furthermore, due to the blurred relationships between the constants after performing substitution, too much information is lost for linear subtree simplification to be effective.

\begin{table}[H]
\centering
\begin{tabular}{|ccccccc|}
\hline
\multicolumn{7}{|c|}{\bfseries MSiMBA}                                                                                                                                                                                                       \\ \hline
\multicolumn{1}{|l|}{\multirow{2}{*}{\bfseries Expr}} & \multicolumn{2}{c|}{\bfseries 2 variables}                                  & \multicolumn{2}{c|}{\bfseries 3 variables}                                  & \multicolumn{2}{c|}{\bfseries 4 variables}             \\ \cline{2-7} 
\multicolumn{1}{|l|}{}                      & \multicolumn{1}{c|}{\bfseries Runtime}   & \multicolumn{1}{c|}{\bfseries \# of Nodes} & \multicolumn{1}{c|}{\bfseries Runtime}   & \multicolumn{1}{c|}{\bfseries \# of Nodes} & \multicolumn{1}{c|}{\bfseries Runtime}   &\bfseries  \# of Nodes \\ \hline
\multicolumn{1}{|c|}{$e_1$}                    & \multicolumn{1}{c|}{0.02 $ms$} & \multicolumn{1}{c|}{3 / 3}       & \multicolumn{1}{c|}{0.04 $ms$} & \multicolumn{1}{c|}{3 / 3}       & \multicolumn{1}{c|}{0.15 $ms$} & 3 / 3       \\ \hline
\multicolumn{1}{|c|}{$e_2$}                    & \multicolumn{1}{c|}{0.01 $ms$} & \multicolumn{1}{c|}{1 / 1}       & \multicolumn{1}{c|}{0.03 $ms$} & \multicolumn{1}{c|}{1 / 1}       & \multicolumn{1}{c|}{0.10 $ms$} & 1 / 1       \\ \hline
\multicolumn{1}{|c|}{$e_3$}                    & \multicolumn{1}{c|}{0.02 $ms$} & \multicolumn{1}{c|}{5 / 5}       & \multicolumn{1}{c|}{0.06 $ms$} & \multicolumn{1}{c|}{5 / 5}       & \multicolumn{1}{c|}{0.15 $ms$} & 5 / 5       \\ \hline
\multicolumn{1}{|c|}{$e_4$}                    & \multicolumn{1}{c|}{0.11 $ms$} & \multicolumn{1}{c|}{7 / 7}       & \multicolumn{1}{c|}{0.06 $ms$} & \multicolumn{1}{c|}{7 / 7}         & \multicolumn{1}{c|}{0.11 $ms$} & 7 / 7       \\ \hline
\multicolumn{1}{|c|}{$e_5$}                    & \multicolumn{1}{c|}{0.03 $ms$} & \multicolumn{1}{c|}{4 / 4}       & \multicolumn{1}{c|}{0.04 $ms$} & \multicolumn{1}{c|}{4 / 4}       & \multicolumn{1}{c|}{0.09 $ms$} & 4 / 4       \\ \hline
\end{tabular}
\caption{Runtime and distance to ground truth of MSiMBA on 1000 MBAs generated for five expressions}\label{tab:our}
\end{table}

Lastly, we see that our algorithm substantially outperforms both ProMBA and GAMBA on semi-linear MBAs. MSiMBA is able to simplify all expressions exactly to their ground truth, and it does so within only fractions of milliseconds.

\subsection{Comparison on self-generated and MBA-Obfuscator datasets}

\begin{table}[H]
\begin{tabular}{|c|cc|cc|cc|}
\hline
\multirow{2}{*}{Dataset} & \multicolumn{2}{c|}{\bfseries ProMBA}                          & \multicolumn{2}{c|}{\bfseries GAMBA}                        & \multicolumn{2}{c|}{\bfseries MSiMBA}                    \\ \cline{2-7} 
                         & \multicolumn{1}{c|}{\bfseries Runtime}        &\bfseries  \# of Nodes    & \multicolumn{1}{c|}{\bfseries Runtime}      &\bfseries  \# of Nodes   & \multicolumn{1}{c|}{\bfseries Runtime}   &\bfseries  \# of Nodes   \\ \hline
$MBA-Obf$                & \multicolumn{1}{c|}{36,571.27 $ms$} & 186.57 / 11.94 & \multicolumn{1}{c|}{743.74 $ms$}  & 39.51 / 11.94 & \multicolumn{1}{c|}{0.13 $ms$} & 12.29 / 11.94 \\ \hline
$Self-generated$           & \multicolumn{1}{c|}{33,345.00 $ms$} & 190.18 / 5     & \multicolumn{1}{c|}{2843.65 $ms$} & 124.12 / 5    & \multicolumn{1}{c|}{0.06 $ms$} & 6.02 / 5      \\ \hline
\end{tabular}
\end{table}

Our algorithm outperforms ProMBA and GAMBA on both datasets. It manages to find an optimal result in almost all cases - and a close to optimal result in the remaining cases.

\section{Conclusion}
This works expands the applicability of algebraic deobfuscation techniques to a broader class of MBAs, for which peer tools cannot simplify. Our extended version of SiMBA is very competitive in runtime, and capable of simplifying all expressions within both self generated and semi-linear versions of public datasets. 

We believe that our approach substantially broadens the types of MBAs that can be effectively deobfuscated. In fact, our algorithm addresses one of the biggest challenges facing MBA deobfuscation: the simplification of linear expressions with nontrivial bitwise constants.

\section{Acknowledgements}
To an anonymous contributor\href{(https://github.com/duk-37)}{(https://github.com/duk-37)}, for contributing his knowledge during countless discussions about MBA obfuscation, and for his time spent collaborating on MBA simplification. To Matteo Favaro, for taking the time to discuss, generate, and simplify challenging nonlinear MBAs.

\bibliographystyle{plain}
\bibliography{sample}

\end{document}